\documentclass[11pt]{article}
\newcommand{\ket}[1]{\left | #1 \right \rangle}
\newcommand{\bra}[1]{\left \langle #1 \right |}

\def\openone{\leavevmode\hbox{\small1\kern-3.8pt\normalsize1}}
\def\RR{{\rm I\kern-.2emR}}

\def\ce{{\cal E}}

\def\crr{{\cal R}}

\newcommand{\proj}[1]{\ket{#1}\!\bra{#1}}

\newcommand{\inner}[2]{ \langle #1 | #2 \rangle}

\newcommand{\beq}{\begin{equation}}
\newcommand{\eeq}{\end{equation}}
\newcommand{\beqa}{\begin{eqnarray}}
\newcommand{\eeqa}{\end{eqnarray}}

\newtheorem{theorem}{Theorem}

\newtheorem{lemma}{Lemma}

\begin{document}
\begin{center}
{\LARGE\bf Illustrating the concept of quantum information\\ }
\bigskip
{\normalsize Richard Jozsa}\\
{\small\it Department of Computer Science, University of
Bristol,\\ Merchant Venturers Building, Bristol BS8 1UB U.K.}
\\[4mm]
\date{today}
\end{center}

\begin{abstract}
Over the past decade quantum information theory has developed into
a vigorous field of research despite the fact that quantum
information, as a precise concept, is undefined. Indeed the very
idea of viewing quantum states as carriers of some kind of
information (albeit unknowable in classical terms), leads
naturally to interesting questions that might otherwise never have
been asked, and corresponding new insights.  We will discuss some
illustrative examples, including a strengthening of the well known
no-cloning theorem leading to a property of permanence for quantum
information, and considerations arising from information
compression that reflect on fundamental issues.
\end{abstract}
\section{Introduction}
Perhaps the most intriguing product of quantum information theory
is the concept of quantum information itself. In the early 1990s
Charles Bennett was one of the first workers to recognise and
promote this new concept, establishing the foundations of a new
subject. Taken as a primary ingredient, quantum information cannot
be defined, but the viewpoint it fosters is richly suggestive,
leading to new interesting questions and modes of interpretation
for some quantum processes. In this paper we will explore a few
examples.

A quantum state $\ket{\psi}$ may be viewed as a carrier of
information in two fundamentally different ways. Firstly
$\ket{\psi}$ may be regarded as carrying the {\em classical}
information of the state identity. As an example, a sender may
prepare one of the two (non-orthogonal) states $\ket{\psi_0}$ and
$\ket{\psi_1}$ to encode the bit values 0 and 1 respectively. Then
the receiver's task is to regain the value of $i$ from
$\ket{\psi_i}$. If $p_{j|i}$ denotes the probability that he
outputs $j$ when the state was $\ket{\psi_i}$ and $q_i$ is the
probability that $\ket{\psi_i}$ was sent, then he may, for
example, apply a procedure that minimises the error probability
$p_{2|1}q_1+p_{1|2}q_2$. In this way the available information in
the quantum state is similar to the result of classical
communication through a noisy channel and it is well known that if
$\inner{\psi_1}{\psi_2}\neq 0$ then the minimum error probability
cannot be zero i.e. the state $\ket{\psi_i}$ cannot be perfectly
identified by any physical process.

In a second way, $\ket{\psi}$ may be viewed as the carrier of
``quantum information'' which although we leave undefined in more
fundamental terms, we intuitively think of as ``the state
itself''. Quantum information is a new concept with no classical
analogue and it is important to distinguish it from the state
identity. For example, given a physical realisation of one of the
two states $\ket{\psi_i}$ above, quantum theory considerably
restricts (in a richly structured way) the allowable manipulations
that we can perform, in contrast to what is possible if we are
given the identity of $i$. Indeed ``being given the quantum state
$\ket{\psi_i}$'' is very different from being given any kind of
classical information and by an analogy of terminology we apply
the name quantum information to describe what we have received.
Stated more formally, we would aim to formulate and interpret
quantum physics in a way that has a concept of information as a
primary fundamental ingredient. Primary fundamental concepts are
{\em ipso facto} undefined (as a definition amounts to a
characterisation in yet more fundamental terms) and they acquire
meaning only afterwards, from the structure of the theory they
support.

As a first example consider the process of quantum teleportation
(cf \cite{telep} for full details): Alice (A) succeeds in
transferring a qubit state to Bob (B) (distantly separated in
space) by sending only two classical bits of information. A and B
also need to share an entangled (EPR) pair which is destroyed in
the process. We would like to think of teleportation as the
transmission of quantum information from A to B. If we accept the
intuitively appealing tenet, that a transfer of information from
sender to receiver must always be mediated by a channel connecting
the two participants, then teleportation appears paradoxical -
only two classical bits were sent so how did the full quantum
information pass from A to B? Looking at the standard spacetime
diagram of the teleportation process (cf figure 1(a) of
\cite{telep}) we see that there is indeed a second (V shaped) path
connecting A to B, which is defined by the two worldlines of the
distributed EPR particle pair. This leads to an intriguing
interpretation (first proposed by Bennett soon after the discovery
of teleportation): in addition to the two bits, the remaining
quantum information must have been propagated {\em backwards in
time} from A to the EPR source and thence forwards in time to
B\footnote{A similar interpretation involving backwards in time
propagation was proposed earlier for the Bennett-Wiesner dense
coding protocol in \cite{bw92} and attributed there to Ben
Schumacher. For further developments of this idea see also
\cite{bentime}.}. Indeed if we insist that information
transmission requires a physical channel then there appears to be
no other possible interpretation of the teleportation process! It
is remarkable that this interpretation is entirely consistent: the
principles of quantum measurement theory imply that the
information sent backwards in time is random and independent of
the teleported state, so long as the two classical bits remain
unknown. Hence the well known classical causal paradoxes of
backwards in time information propagation are neatly circumvented.
This analysis, inspired by our informational point of view, also
reveals a new significance for entanglement in quantum theory
(beyond the traditional issues of non-local correlations of
 measurement outcomes): entanglement can be viewed as
providing a channel for the transmission of quantum information.

In the following sections we will discuss two further issues in
which an informational point of view leads to interesting
considerations. Firstly we will revisit the quantum no-cloning
theorem \cite{WZ} and prove a new stronger form of this result.
Together with the Pati-Braunstein no-deleting principle \cite{PB}
this will lead to a property of ``permanence'' for quantum
information. Secondly we will discuss the concept of information
compression. In classical information theory this provides one of
the clearest approaches to the concept of information. By
mimicking this theory in a quantum context we obtain some
surprising relationships between the concept of information and
the geometry of Hilbert space (i.e. the basis of the conventional
formulation of quantum theory).

\section{A stronger no-cloning theorem}
It is well known that (non-orthogonal) pure quantum states cannot
be cloned \cite{WZ} i.e. if $\{ \ket{\psi_i}\}$ is a set of pure
states containing at least one non-orthogonal pair then no
physical operation can achieve the transformation $ \ket{\psi_i}
\longrightarrow \ket{\psi_i}\ket{\psi_i}$. Although the
impossibility of cloning in quantum theory can be attributed to
the fact that such a process is non-unitary or non-linear, from an
informational point of view we can intuitively understand it by
saying that two copies of a quantum state embody strictly more
``information'' than is available in just one copy, so cloning
must be impossible. Extending this particular line of thought, it
is then natural to go on to ask: what additional (quantum)
information is needed to supplement one copy $\ket{\psi_i}$ in
order to be able to produce two copies $\ket{\psi_i}\ket{\psi_i}$?
For classical information, no supplementary information at all is
needed and one might guess that as the set $\{ \ket{\psi_i}\}$
becomes ``more classical'' the necessary supplementary information
should decrease in some suitable way. However we prove below that
this is not the case: we show (for mutually non-orthogonal states)
that the supplementary information must always be as large as it
can possibly be i.e. the second copy $\ket{\psi_i}$ can always be
generated from the supplementary information alone, independently
of the first (given) copy. Thus in effect, cloning of
$\ket{\psi_i}$ is possible only if the second copy is fully
provided as an additional input.

We now give a precise formulation of the main result in this
section. By a physical operation we will mean a trace preserving
completely positive map. Note that this definition excludes the
collapse of wavefunction in a quantum measurement, as a valid
physical process. (This will be relevant to our later discussion
of the no-deleting principle.) By an abuse of notation for pure
states we will write $\proj{\psi}\otimes \rho$ as just
$\ket{\psi}\otimes \rho$ and sometimes also omit the tensor
product symbol, writing $\proj{\psi}\otimes \proj{\psi}$ as
$\ket{\psi}\ket{\psi}$.

{\theorem \label{thm1} Let $\{ \ket{\psi_i}\}$ be any finite set
of pure states containing no orthogonal pairs of states. Let $\{
\rho_i \}$ be any other set of (generally mixed) states indexed by
the same labels. Then there is a physical operation \[
\ket{\psi_i}\otimes \rho_i \longrightarrow
\ket{\psi_i}\ket{\psi_i} \] if and only if there is a physical
operation \[ \rho_i \longrightarrow \ket{\psi_i} \] i.e. the full
information of the clone must already be provided in the ancilla
state $\rho_i$ alone.}

\noindent {\bf Remark}\cite{errors}. If the set $\{ \ket{\psi_i}
\}$ contains some orthogonal pairs then the unassisted clonability
of orthogonal states spoils the simplicity of the statement of
theorem 1. As an example consider
\[ \begin{array}{ll} \ket{\psi_1}=\ket{0} &
\ket{\alpha_1}=\ket{a}  \\   \ket{\psi_2}=\ket{1} &
\ket{\alpha_2}=\ket{a} \\ \ket{\psi_3} = \frac{1}{\sqrt{2}}
(\ket{0}+\ket{1}) & \ket{\alpha_3}=\ket{b}
\end{array}
\]
where $\ket{a}$ and $\ket{b}$ are orthogonal. Then clearly
$\ket{\psi_i}\ket{\alpha_i} \rightarrow \ket{\psi_i}\ket{\psi_i}$
is possible (as $\{ \ket{\psi_i}\ket{\alpha_i} \}$ is an
orthonormal set) but $\ket{\alpha_i}\rightarrow \ket{\psi_i}$ is
not possible (as $\ket{\alpha_1}=\ket{\alpha_2}$ but
$\ket{\psi_1}\neq \ket{\psi_2}$). Indeed the $\ket{\alpha_i}$
states here provide reliable distinguishability of $i$ values
exactly when this is not already provided by the $\ket{\psi_i}$'s
themselves.

To prove the theorem we will use the following lemma which is
proved as lemma 1 of \cite{JS}. {\lemma \label{lemma1} Let $\{
\ket{\alpha_i} \}$ and $\{ \ket{\beta_i} \}$ be two sets of pure
states (indexed by the same labels). Then the two sets have equal
matrices of inner products (i.e.
$\inner{\alpha_i}{\alpha_j}=\inner{\beta_i}{\beta_j}$ for all $i$
and $j$) if and only if the two sets are unitarily equivalent
(i.e. there exists a unitary operation $U$ on the direct sum of
the state spaces of the two sets with
$U\ket{\alpha_i}=\ket{\beta_i}$ for all $i$). }

\noindent {\bf Proof of theorem \ref{thm1}} \,\, Suppose that
there is a physical operation $\rho_i \rightarrow \ket{\psi_i}$.
Then clearly $\ket{\psi_i}\otimes \rho_i \rightarrow
\ket{\psi_i}\ket{\psi_i}$ is allowed.

Conversely suppose that there is a physical operation
\begin{equation} \label{cl}
\ket{\psi_i}\otimes \rho_i \longrightarrow
\ket{\psi_i}\ket{\psi_i}. \end{equation} Consider first the case
that $\rho_i$ are pure states, $\ket{\alpha_i}$ say. The physical
operation eq. (\ref{cl}) may be expressed as a unitary operation
if we include an environment space, initially in a fixed state
denoted $\ket{A}$. For clarity we include an extra register,
initially in a fixed state $\ket{0}$, that is to receive the clone
of $\ket{\psi_i}$. Then eq. (\ref{cl}) may be written as a {\em
unitary} transformation \[
\ket{\psi_i}\ket{0}\ket{\alpha_i}\ket{A} \longrightarrow
\ket{\psi_i}\ket{\psi_i}\ket{C_i} \] where $\ket{C_i}$ (generally
depending on $i$) is the output state of the two registers that
initially contained $\ket{\alpha_i}\ket{A}$. Hence by unitarity,
the two sets $\{ \ket{\psi_i}\ket{\alpha_i} \}$ and $\{
\ket{\psi_i}\ket{\psi_i}\ket{C_i} \}$ have equal matrices of inner
products and then, so do the sets $\{ \ket{\alpha_i} \} $ and $\{
\ket{\psi_i} \ket{C_i} \}$ (by a simple cancellation of
$\inner{\psi_i}{\psi_j}$ from the two initial matrices). Thus by
lemma \ref{lemma1} these two sets are unitarily equivalent so
$\ket{\psi_i}$ can be generated from $\ket{\alpha_i}$ alone (by
applying the unitary equivalence and discarding the $\ket{C_i}$
register).

If $\rho_i$ are mixed we express them as probabilistic mixtures of
pure states \[ \rho_i = \sum_k p^{(i)}_k \proj{\alpha^{(i)}_k} \]
(where all $p_k^{(i)}$ are non-zero). Then a physical operation
achieves
\[ \ket{\psi_i}\otimes \rho_i \longrightarrow \ket{\psi_i}
\ket{\psi_i} \hspace{1cm} \mbox{for all $i$}
\] if and only if it achieves \begin{equation} \label{eq1}
 \ket{\psi_i}\otimes \ket{\alpha^{(i)}_k}
\longrightarrow \ket{\psi_i} \ket{\psi_i} \hspace{1cm} \mbox{for
all $i$ and $k$.} \end{equation}  By the pure state analysis
above, a physical operation effecting eq. (\ref{eq1}) exists if
and only if there is a physical operation effecting \[
\ket{\alpha^{(i)}_k} \longrightarrow \ket{\psi_i} \hspace{1cm}
\mbox{for all $i$ and $k$} \] and then we get $\rho_i
\longrightarrow \ket{\psi_i}$ too. $QED$.

In the particular case of cloning assisted by {\em classical}
information (i.e. the states $\rho_i$ are required to be mutually
commuting) we deduce that this supplementary data must contain the
full identity of the states as classical information. Indeed if
the $\rho_i$ are classical then they can be copied any number of
times so if we can make one clone of $\ket{\psi_i}$ from $\rho_i$,
we can make arbitrarily many clones and hence determine the the
identity of $\ket{\psi_i}$ i.e. the classical information of the
label $i$ must be contained in the supplementary classical
information.

The proof of theorem \ref{thm1} is easily adapted to prove the
following generalisation. Let $\{ \ket{\psi_i}\}$ be any finite
set of pure states containing no orthogonal pairs of states. Let
$\ket{\psi_i}^{\otimes n}$ denote the state of $n$ copies of
$\ket{\psi_i}$.  Then there is a physical operation \[
\ket{\psi_i}^{\otimes n} \otimes \rho_i \longrightarrow
\ket{\psi_i}^{\otimes (n+1)} \] if and only if there is a physical
operation
\[ \rho_i \longrightarrow \ket{\psi_i}. \] Curiously, the
increasing information contained in $n$ copies of $\ket{\psi_i}$
(as $n$ increases) can never be used to assist in the creation of
even a single extra copy.

\section{No deleting}
Our techniques may also be used to give a simple proof of the
Pati-Braunstein no-deleting principle \cite{PB} for sets $\{
\ket{\psi_i} \}$ that contain no orthogonal pairs. The issue here
is the following. Suppose we have two copies
$\ket{\psi_i}\ket{\psi_i}$ of a state and we wish to delete one
copy by a physical operation:
\begin{equation} \label{nd1} \ket{\psi_i}\ket{\psi_i} \longrightarrow
\ket{\psi_i}\ket{0} \end{equation} (where $\ket{0}$ is any fixed
state of the second register). As before, any such physical
operation may be expressed as a unitary operation if we include an
environment space, initially in a fixed state $\ket{A}$ say. Then
eq. (\ref{nd1}) is equivalent to the unitary transformation
\begin{equation} \label{nd2} \ket{\psi_i}\ket{\psi_i}\ket{A}
\longrightarrow \ket{\psi_i}\ket{0}\ket{A_i} \end{equation} where
the final state $\ket{A_i}$ of the environment may depend on
$\ket{\psi_i}$ in general. One way of achieving this is to simply
swap (a constant) part of the environment into the second register
but then the second copy of $\ket{\psi_i}$ remains in existence
(albeit in the environment now). The no-deleting principle states
that the second copy of $\ket{\psi_i}$ can {\em never} be
``deleted'' in the sense that $\ket{\psi_i}$ can {\em always} be
resurrected from $\ket{A_i}$. Note however, that if wavefunction
collapse is also allowed as a valid physical process then deletion
is possible. (We perform a complete measurement on $\ket{\psi_i}$
and rotate the seen post-measurement state to $\ket{0}$ by a
unitary transformation that depends on the measurement outcome.)

To see the no-deleting principle with our methods, note that the
unitarity of eq. (\ref{nd2}) implies that the sets $\{
\ket{\psi_i}\ket{\psi_i}\ket{A} \}$ and $\{
\ket{\psi_i}\ket{0}\ket{A_i} \}$ have equal matrices of inner
products and then as before, so do the sets $\{ \ket{\psi_i} \}$
and $\{ \ket{A_i} \}$. Thus lemma \ref{lemma1} states that these
sets are unitarily equivalent, which is just the no-deleting
principle.

\section{Permanence of quantum information}

Deleting and cloning have a common feature: in cloning we saw that
the existence of the first copy $\ket{\psi_i}$ provided no
assistance in constructing the second copy from the supplementary
information. Similarly for deletion, the existence of the first
copy provides no assistance in deleting the second copy -- in
effect the only way to delete the second copy is to transform it
out into the environment (i.e. $\ket{0}\ket{A_i} $ in eq.
(\ref{nd2}) is a unitary transform of $\ket{\psi_i}\ket{A}$ alone)
and this again makes no use of the first copy. Considering
no-cloning and no-deleting together (and excluding wavefunction
collapse as a valid physical process) we see that quantum
information (of non-orthogonal states) has a quality of
``permanence'': creation of copies can only be achieved by
importing the information from some other part of the world where
it had {\em already existed}; destruction (deletion of a copy) can
only be achieved by exporting the information out to some other
part of the world where it must {\em continue to exist}. This
property is different from the {\em preservation} of information
by any reversible dynamics. For example the classical reversible
C-NOT operation can imprint copies of a bit $b$ into a standard
state via $\ket{b}\ket{0} \rightarrow \ket{b}\ket{b}$ and also
delete copies via $\ket{b}\ket{b} \rightarrow \ket{b}\ket{0}$ but
in both cases the first copy is used in an essential way in the
process and the information content of one copy is the same as
that of two copies. In contrast, in the quantum (non-orthogonal)
case, copying and deleting can only occur independently of the
first copy and then reversibility of dynamics implies that the
information of the second copy must have already separately
existed in the environment (for cloning) or continue to exist
separately in the environment (for deletion). But in any
reasonable intuitive sense, $\ket{\psi_i}\ket{\psi_i}$ does not
have double the information content of $\ket{\psi_i}$ (and
similarly $\ket{\psi_i}^{\otimes n}$ cannot have $n$ times the
information content, as $n$ is unbounded). One might interpret
this as an overlap of information content of the two copies and
then theorem \ref{thm1} implies that this common part cannot be
duplicated from within a single copy and merely extended, to give
the second copy.

\section{Information compression and Hilbert space geometry}

So far our discussion of quantum information has been qualitative.
However it would be interesting to develop a quantitative theory
of this concept, being able to say that one quantum system has
more quantum information than another, and we would like to have
corresponding dynamical laws for the manipulation of quantum
information. In classical information theory there exists a well
defined quantitative notion of information and as a first attempt
we can consider importing it into a quantum context.

In Shannon's classical information theory we begin with a
classical information source which is defined by a prior
probability distribution $ \{ p_i \}$ of signals $s_i$. The
information content is then quantified by the Shannon entropy
$H(p_i)=-\sum_i p_i \log_2 p_i$ bits. This definition has a
compelling physical interpretation: it characterises the minimal
resources ($H$ bits per signal) that are necessary and sufficient
to faithfully represent the source (in a suitable asymptotic sense
\cite{CT} that we need not elaborate here). To approach the
concept of quantum information, a natural avenue is to mimic this
very successful classical theory in a quantum context. Thus we
introduce a quantum source, characterised by a prior probability
distribution $\{ p_i \}$ of quantum signals $\ket{\alpha_i}$ and
define its quantum information content to be $S$, the least number
of qubits that are necessary and sufficient to faithfully
represent the source (in an asymptotic sense that naturally
generalises the classical case). Following Schumacher, one may
prove \cite{compression} that $S$ is then the von Neumann entropy
$S(\rho)$ of the source density matrix $\rho=\sum_i p_i
\proj{\alpha_i}$, establishing a fundamental role for von Neumann
entropy in quantum information theory.

This notion of quantum information, while interesting, is perhaps
not entirely satisfactory in that it still involves an essentially
classical ingredient {\em viz.} the prior classical probability
distribution. However in this context it should be pointed out
that that there is an unexpected and remarkable harmony in such
classical mixing of quantum information: if $\ce_1 =\{
\ket{\alpha_i}; p_i \}$ and $\ce_2 = \{ \ket{\beta_j}; q_j \}$ are
two quantum sources with the same density matrix $\sum_i p_i
\proj{\alpha_i} = \sum_j q_j \proj{\beta_j}$ then $\ce_1$ and
$\ce_2$ are entirely indistinguishable by any physical process.
The quantum information of the $\ket{\alpha_i}$'s
probabilistically mixed by $p_i$ is exactly the same as the
quantum information of the $\ket{\beta_j}$'s mixed by $q_j$; no
trace of the component states remains in the mixture! This
indistinguishability can be seen to be related to various other
consistency requirements of a physical theory such as the no
superluminal signalling principle\cite{HJW}.

A second difficulty with the proposal of identifying the quantum
information content of a source with its von Neumann entropy $S$
is the fact that very different sources can have the same entropy
yet some look ``more quantum'' than others! This was realised soon
after the appearance of Schumacher's compression theorem and
discussed by Bennett and other participants during the first
Elsag-Bailey quantum computing workshop at the ISI in Torino in
July 1993. It was suggested that a quantum source might be
decomposable into a classical and a quantum part, with the von
Neumann entropy quantifying both parts together. Then we would
seek to separate out a maximal classical part and quantify the
quantum part alone\cite{oldrevext}. To illustrate the situation
consider a source which emits one of two orthogonal states
$\ket{\psi_0}$ and $\ket{\psi_1}$ with equal prior probabilities
of $\frac{1}{2}$. As the states can be reliably identified by a
measurement, this source can be represented entirely in classical
terms with $S=1$ classical bit per signal. Suppose now that the
states are not quite orthogonal e.g. $|\inner{\psi_0}{\psi_1}| =
10^{-9}$. The von Neumann entropy is still very close to 1 and we
ask: is this source ``almost classical''? i.e. can we extract
approximately 1 classical bit of information leaving behind a very
small amount of quantum information (e.g. almost parallel states)
in such a way that the signals can still be faithfully
reconstructed from the classical and quantum information parts?
This question was settled only recently \cite{revext}, in the
negative: Let $\ce$ be a quantum source whose signal states do not
lie in a family of orthogonal subspaces. Then $\ce$ can be
faithfully compressed to $\alpha$ qubits per signal plus any
number of classical bits per signal if and only if $\alpha$ is at
least as large as the von Neumann entropy $S$ i.e. it is
generically impossible to separate a source non-trivially into a
classical and a quantum part, and the classical representation of
exactly orthogonal states is therefore a singular feature of
infinite precision.

Thus we need to look at more subtle properties of quantum
compression to distinguish sources with equal entropy. Following a
further suggestion of Bennett, we can study features of so-called
{\em visible} quantum compression. In this scenario the source is
described by giving the {\em classical} information of the emitted
signal state's identity (rather than the quantum state itself, as
quantum information). Our task again is to faithfully represent
the signal states with minimal resources and now we have more
possibilities for coding. As an example consider a source of four
signal states, with equal prior probabilities of $\frac{1}{4}$ and
having von Neumann entropy 1. In visible compression we can
represent this source with 2 classical bits per signal and no
qubits (as there are four equiprobable possibilities) or with 1
qubit per signal and no classical bits (by creating the signal
states and performing Schumacher's quantum compression on them).
In between these two extremes there is a trade-off curve $q(c)$:
if we have $c$ classical bits per signal (with $0\leq c\leq 2$)
then $q(c)$ is the least number of qubits per signal that is
sufficient to represent the source (so the above gives $q(0)=1$
and $q(2)=0$). Thus instead of trying to extract classical
information from a quantum source, we {\em start} by giving a
fully classical description of the source and consider the
trade-off involved in coding the source back into quantum terms.
An extensive study of this trade-off curve is given in
\cite{tradeoff} and it is found that the shape of the curve does
indeed distinguish different sources with the same entropy.

Returning now to information compression and the insight it may
provide into the notion of information, it is interesting to ask
why compression is possible at all. Evidently some kind of
redundancy in the raw signals is being eliminated. For a classical
source it is well known that non-trivial compression is possible
if and only if the prior probability distribution is not uniform.
For example consider the case of two signals with unequal
probabilities. In that case we already have some prior knowledge
of the signal before it is received, in the sense that we can
guess the signal (choosing the more probable one) and be correct
more often than not. In this sense part of the signal (if sent in
full) is redundant.

The quantum situation is considerably more subtle. For any quantum
source $\{ \ket{\alpha_i}; p_i \}$ we have $S(\rho) \leq H(p_i)$
with equality if and only if the signals are all mutually
orthogonal, suggesting that there is a quantum redundancy
associated specifically with non-orthogonality. For example
consider two qubit signals $\ket{\psi_1}$ and $\ket{\psi_2}$ at
$45^\circ$ with equal prior probabilities of $\frac{1}{2}$. Then
$H(p_i)$ is 1 bit but $S(\rho)$ is 0.601 qubits. Moreover
$S(\rho)$ decreases monotonically from 1 to 0 as the overlap
$|\inner{\psi_1}{\psi_2}|^2$ is increased from 0 to 1.

The interpretation of non-orthogonality is one of the enigmas of
quantum theory. Conventionally the overlap
$|\inner{\psi_1}{\psi_2}|^2$ provides a measure of the
non-distinguishability of $\ket{\psi_1}$ and $\ket{\psi_2}$ and
this is reflected in the properties of the von Neumann entropy
above {\em viz.} an increasing redundancy of quantum information
with increasing overlap. Thus if $\ce_1= \{ \ket{\alpha_i};
\frac{1}{n} \}$ and $\ce_2=\{ \ket{\beta_i};\frac{1}{n} \}$ are
two quantum sources with $n$ states each (having all prior
probabilities equal, for simplicity) and with larger overlaps
$|\inner{\beta_i}{\beta_j}|^2 > |\inner{\alpha_i}{\alpha_j}|^2$
for each pair in $\ce_2$ compared to the corresponding pair in
$\ce_1$, then we would expect a decrease of information content in
passing from $\ce_1$ to $\ce_2$. While this is true for $n=2$ it
can be shown to fail generically\cite{JS} for $n=3$ and higher
i.e. it is generically possible to increase the von Neumann
entropy of a source while increasing the overlaps of every pair of
signal states! Evidently the quantum information content depends
on more subtle structural properties of the geometry of the
signals in the Hilbert space, beyond just the pairwise overlaps.

The relationship of quantum information to the geometry of the
Hilbert space is largely unstudied but for $n=3$ we can say a
little more\cite{JS}. The von Neumann entropy $S$ of three
equiprobable states $\ket{\psi_1}$, $\ket{\psi_2}$, $\ket{\psi_3}$
is a function of four independent real parameters: the three
overlaps $a_{12}=|\inner{\psi_1}{\psi_2}|^2$,
$a_{23}=|\inner{\psi_2}{\psi_3}|^2$,
$a_{31}=|\inner{\psi_3}{\psi_1}|^2$ and the phase $\xi$ of the
triple product $\Upsilon =
\inner{\psi_1}{\psi_2}\inner{\psi_2}{\psi_3}\inner{\psi_3}{\psi_1}$
(noting that the squared modulus of $\Upsilon$ is the dependent
quantity $a_{12}a_{23}a_{31}$). Then keeping $a_{12}$, $a_{23}$
and $a_{31}$ fixed we can vary $\xi$ and it can be shown \cite{JS}
that $S$ is actually a monotonically decreasing function of $\cos
\xi$ (or $\crr e (\Upsilon)$). Despite this clean relationship, we
still lack an intuitive understanding of why increasing the phase
$\xi$ allows increased compressibility.

In this paper we have promoted a viewpoint that attempts to place
a notion of information at a primary fundamental level in the
formulation of quantum physics. In the spirit of Landauer's slogan
``Information is physical!'' we would declare ``Physics is
informational!'' Physical theories have traditionally been
formulated in conceptual and mathematical terms that are, at root,
{\em geometrical}. As such, they have an intuitive accessibility
which has facilitated many developments (for example, the powerful
guiding principles of symmetry and co-ordinate invariance in the
construction of Lagrangians and field equations). Similarly the
concept of information has an intuitive basis although not
geometrical (and evidently having a complicated relation to the
geometry of state space). Hence it offers a potentially new
perspective on quantum physics with its own characteristic flavour
of guiding principles. For example we might adopt the principle
that any prospective physical theory should not allow the
efficient solution of an NP complete problem. This principle
greatly restricts the form of the theory yet remarkably, it
appears to hold in the established formalisms of classical and
quantum physics, which developed from entirely different
perspectives. Although an informational and geometrical
formulation of a given physical theory would be mathematically
equivalent, both points of view are valuable for further
developments: natural generalisations that these respective
formalisms suggest would be quite different and no longer
equivalent as theories.

\bigskip

\noindent {\Large\bf Acknowledgements}

\noindent RJ is supported by the U.K. Engineering and Physical
Sciences Research Council.

\end{document}